\documentclass[a4paper,11pt]{article}
\usepackage{cite}
\usepackage{graphicx}
\usepackage{float}
\usepackage{wrapfig}
\usepackage{authblk}

\begin{document}

\title{Projectile motion without calculus}
\author{Joseph A Rizcallah\\joeriz68@gmail.com}
\affil{School of Education, Lebanese University, Beirut, Lebanon}
\date{}
\maketitle

\begin{abstract}
Projectile motion is a constant theme in introductory-physics courses. It is often used to illustrate the application of differential and integral calculus. While most of the problems used for this purpose, such as maximizing the range, are kept at a fairly elementary level, some, such as determining the safe domain, involve not so elementary techniques, which can hardly be assumed of the targeted audience. In the literature, several attempts have been undertaken to avoid calculus altogether and keep the exposition entirely within the realm of algebra and/or geometry. In this paper, we propose yet another non-calculus approach which uses the projectile's travel times to shed new light on these problems and provide instructors with an alternate method to address them with their students.
\end{abstract}

\section{Introduction}\label{intro}
Projectile motion is a common topic to all introductory mechanics courses, where students come to apply the newly acquired methods of kinematics to explore motion in a uniform gravitational field. In most introductory textbooks, algebra-based as well as calculus-based~\cite{gian,giam,young,tip,ser}, the equations of projectile motion are derived by making use of the superposition of uniform and uniformly accelerated rectilinear motions in the horizontal and vertical directions respectively. In this respect, the algebra-based courses are no different from their calculus counterparts, which seem to standout mainly by offering the student some calculus-oriented problems calling for the application of derivatives, particularly to finding minima and maxima~\cite{young,tip,ser}. 

No doubt, calculus is an indispensable tool for any serious physics student and the aforementioned problems do the students a great service in providing them with a familiar context within which the elementary notions of calculus can be demonstrated and honed. Their downside, however, is their inaccessibility virtually to all students in algebra-based courses. To bridge this gap and enrich the students' experience in the latter courses, several authors have developed purely algebraic and/or geometric approaches to tackle these problems. Apart from the primary goal they are envisaged to serve, these alternate approaches are often far more elegant and insightful than the straightforward calculus-trodden path, that they are welcome by instructors and appreciated by students in algebra-based and calculus-based courses alike. Below is a review of some of these approaches.

In the elegant method of~\cite{pal}, the dot and cross product of velocity vectors are used to solve the maximum range problem. Although it does not involve calculus, this method may prove inappropriate for an introductory course, as the students may not be familiar with these operations from vector algebra. In this respect, the method of completing the squares, suggested in~\cite{bose}, seems to be more appropriate for introductory algebra-based courses (see also~\cite{bace}). Another, more geometric, approach is suggested in~\cite{ganci}. Note, however, that all these methods were developed to solve the level range problem and are not readily applicable to the general range problem, e.g. on an inclined plane. A calculus-heavy solution of the latter, in all its generality, can be found in~\cite{bajc}.  

In the paper, we present a different approach to the maximum range problem. We use displacement triangles (referred to as diagrams in~\cite{noll}) together with travel times to gain a different perspective on the problem. Throughout the paper, we assume a uniform gravitational field and neglect air resistance.   
\section{Level range}
In this section we consider the maximum range problem on a level ground. Its well-known non-calculus solution can be found in any introductory text~\cite{gian,giam,young,tip,ser}, and as shown in~\cite{moh}, the problem can be solved even without recourse to trigonometry. Nevertheless, we choose to include this problem here to introduce our notation and illustrate our approach within a familiar setting. 

The displacement $\vec{r}$ of the projectile at any instant of time $t$ is given by $\vec{r}=1/2\vec{g}t^2+\vec{v}t$, where $\vec{g}$ and $\vec{v}$ denote the acceleration due to gravity and the projectile's initial velocity respectively. Graphically, this vector sum is represented by a displacement triangle with one vertical side. On a level ground, the range occurs when $\vec{r}\cdot\vec{g}=0$. So the corresponding displacement triangle is right of hypotenuse $vt$. We thus have
\begin{equation}
\label{rangecond}
\frac{1}{4}g^2t^4-v^2t^2+R^2=0,
\end{equation}
where $R$ denotes the range, i.e. the value of $r$ when $\vec{r}\cdot\vec{g}=0$ . Viewed as an equation in $t$, with given parameters $R$ and $v$, (\ref{rangecond}) admits two positive roots, $t_1$ and $t_2$, which satisfy the following (Vieta's) relations
\begin{equation}
\label{vieta1}
t_1^2t_2^2=4\frac{R^2}{g^2},
\end{equation}
and
\begin{equation}
\label{vieta2}
t_1^2+t_2^2=4\frac{v^2}{g^2}.
\end{equation}
Hence, for a given pair $R$ and $v$ two travel times are generally possible. Clearly, the different travel times, $t_1$ and $t_2$, correspond to different launch angles, $\theta_1$ and $\theta_2$, correlated with $t_1$ and $t_2$ through $t_i=2v\sin{\theta_i}/g$, for $i = 1, 2$.

Equation (\ref{vieta2}) is truly remarkable. It tells us that the sum of the squares of the travel times is independent of the range $R$. In particular, using (\ref{vieta2}) together with the above relations between $t_i$ and $\theta_i$ yields $\sin^2{\theta_1}+\sin^2{\theta_2}=1$, which implies the well-known complementarity of the launch angles, i.e. $\theta_1 + \theta_2 = \pi/2$, for a given pair $R$ and $v$. Besides, combining (\ref{vieta1}) and (\ref{vieta2}), one obtains
\begin{equation}
\label{range}
R = \frac{v^2}{g} - \frac{1}{4}g(t_1-t_2)^2,
\end{equation}
\begin{wrapfigure}{r}{0.42\textwidth}
\centering
\includegraphics[scale=0.35]{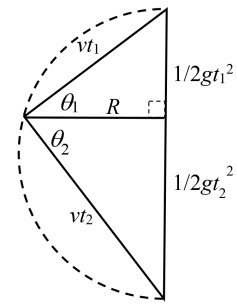}
\caption{The displacement triangles (one flipped upside down) of two projectiles with the same range $R$ but different launch angles.}
\label{fig1}
\end{wrapfigure}
from which it readily appears that the range admits a maximum when $t_1 = t_2$. By the aforementioned complementarity, this implies that the maximum range is attained at $\theta_1 = \theta_2 = \pi/4$.

Equations (\ref{vieta1}) and (\ref{vieta2}) allow a nice geometric interpretation.  Figure~\ref{fig1} shows the two displacement triangles, for a range $R$ and of travel times $t_1$ and $t_2$, with the triangle corresponding to $t_2$ flipped upside down to form a bigger triangle with that corresponding to $t_1$. The complementarity condition is equivalent to the big triangle being right. The latter is established using (\ref{vieta2}) twice as follows: $1/2g(t_1^2+t_2^2) = 2v^2/g=\sqrt{v^2t_1^2+v^2t_2^2}$, so by the Pythagorean theorem the side $1/2g(t_1^2+t_2^2)$ is the hypotenuse and the big triangle is therefore right-angled at the opposite vertex, i.e. $\theta_1 + \theta_2 = \pi/2$, independently of $R$.

Moreover, using (\ref{vieta1}) we readily see that the area $A=1/2v^2t_1t_2$ of this triangle and the range $R$ are related through the proportionality $R=(g/v^2)A$. Now, the independence of the big triangle's hypotenuse and right angle of $R$, means that, as $R$ varies, the triangle's right-angled vertex describes a circular arc (outlined in figure~\ref{fig1}) of diameter $2v^2/g$, i.e. congruent to the hypotenuse. It now becomes clear, that the area of the big triangle, and consequently the range, attains a maximum when the right-angled vertex is at its farthest from the diameter. This, of course, happens when this vertex is a radius $v^2/g$ away from the diameter and the big triangle is right isosceles with equal legs $vt_1 = vt_2$, or equivalently $\theta_1 = \theta_2$. 

In passing, let us note that the projectile's travel time is proportional to the length of the chord drawn from the vertical diameter's ends. In particular, the longest travel time ($2v/g$), for a vertically shot projectile, corresponds to the circle's vertical diameter, while the shortest travel time (zero), for a horizontally shot projectile, corresponds to a tangent to the circle drawn at that diameter. 
\section{Inclined range and safe domain} 
We now turn to the problem of maximum range on a slope of elevation $\alpha$. The vector equation $\vec{r}=1/2\vec{g}t^2+\vec{v}t$ still holds, but the condition for the range now reads $\vec{r}\cdot\vec{g}=rg\cos\varphi$, where $\varphi = \pi/2 +\alpha$. Rearranging terms in the last equation and squaring, we find   
\begin{equation}
\label{rangecond1}
\frac{1}{4}g^2t^4-(v^2+Rg\cos\varphi)t^2+R^2=0,
\end{equation}
where, as before, $R$ denotes the projectile's range. The positive roots $t_1$ and $t_2$ of this equation, corresponding to launch angles $\theta_1$ and $\theta_2$ with the slope, have a product as in (\ref{vieta1}) and a sum of squares given by
\begin{equation}
\label{vieta3}
t_1^2+t_2^2=4\frac{v^2+Rg\cos\varphi}{g^2},
\end{equation}
which upon using (\ref{vieta1}) can be recast into 
\begin{equation}
\label{vieta4}
t_1^2+t_2^2-2t_1t_2\cos\varphi=4\frac{v^2}{g^2}.
\end{equation}
\begin{wrapfigure}{r}{0.37\textwidth}
\begin{center}
\includegraphics[scale=0.33]{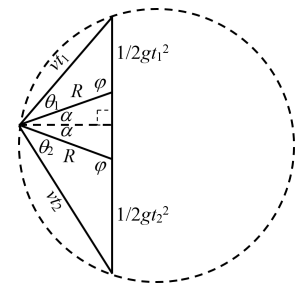}
\end{center}
\caption{The displacement triangles for the inclined range problem. The big triangle's vertical side and opposite angle are independent of $R$.}
\label{fig2}
\end{wrapfigure}

We are interested in the geometric interpretation of (\ref{vieta4}). To this end, consider the displacement triangles associated with $t_1$ and $t_2$, with the latter flipped upside down, so as to form a wedge of angle $2\alpha$ as illustrated in figure~\ref{fig2}. Our goal is to show that, in the big triangle of sides $vt_1$ and $vt_2$, the vertical side, let's call it $a$, is independent of the range $R$ and that the opposite angle, i.e. $\theta_1 + \theta_2 + 2\alpha$ , is equal to $\varphi$. From figure~\ref{fig2}, we see that $a=1/2g(t_1^2+t_2^2)-2R\cos\varphi$. Squaring this and using (\ref{vieta3}), after some algebra, we obtain $a^2= 4v^4/g^2$, which proves the independence of $a$ and $R$. Moreover, by (\ref{vieta4}) we have $a^2= v^2(t_1^2+t_2^2-2t_1t_2\cos\varphi)$, which implies that the angle included between the sides $vt_1$ and $vt_2$ is $\varphi$, leading to $\theta_1 + \theta_2 = \pi/2 - \alpha$, a generalization of the complementarity condition to the present case of inclined range.

It is now easy to see that, as the launch angle varies, the vertex of the big triangle moves on a circle. This is so, because the angle at the vertex and the opposite side are both constant. It is straightforward to show that the radius of this circle equals $v^2/(g\cos\alpha)$. On the other hand, from (\ref{vieta1}) it follows that the area $A=1/2v^2t_1t_2\sin\varphi$ of the big triangle is proportional to the range, i.e. $A=(v^2/g)R\sin\varphi$. Therefore, the maximum range is attained when the vertex is farthest away from $a$. This occurs at the perpendicular bisector of $a$, i.e. when $vt_1 = vt_2$, with the corresponding launch angles being equal $\theta_1 = \theta_2 = \pi/4 - \alpha/2$. 

Note that figure~\ref{fig2} depicts the case of $\alpha > 0$. However, one can easily convince oneself that the same construction works for $\alpha < 0$; all one must do is place the vertex of the big triangle on the complementary arc, i.e. the large arc in figure~\ref{fig2}. It is worth noting though, that in this case the projectile's travel time (proportional to the length of a chord) may exceed that of the vertically shot projectile (proportional to $a$).  In particular the longest travel time, corresponding to the circle's diameter, is $2v/(g\cos\alpha)$ and occurs when the projectile is shot perpendicularly to the slope.

Finally, combining (\ref{vieta1}) and (\ref{vieta3}), one has 
$$
R =\frac{1}{1+\sin\alpha}\left[ \frac{v^2}{g} - \frac{1}{4}g(t_1-t_2)^2\right],
$$
from which we find
\begin{equation}
\label{maxrange}
R_m =\frac{v^2}{g}\frac{1}{1+\sin\alpha}=\frac{v^2}{g}\frac{1}{1+\cos(\pi/2 - \alpha)},
\end{equation}
\begin{wrapfigure}{r}{0.42\textwidth}
\begin{center}
\includegraphics[scale=0.35]{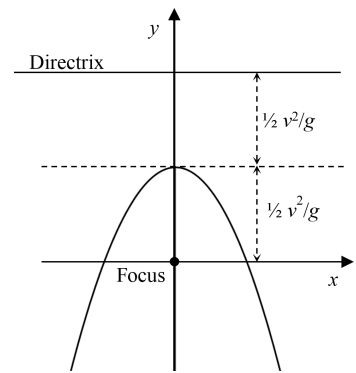}
\end{center}
\caption{The enveloping parabola together with its directrix and focus at the launching point, taken as the origin.}
\label{fig3}
\end{wrapfigure}
for the maximum range $R_m$. Alternatively, (\ref{maxrange}) can be arrived at using the geometry of figure~\ref{fig2} in the special case when the bisector of the wedge coincides with the perpendicular bisector of the chord $a$.

Equation (\ref{maxrange}) defines a parabola (the so called {\it enveloping parabola}) in polar coordinates, $R_m$ and $\alpha$, with a focus at the launching point and a horizontal directrix a distance $v^2/g$ above it (see figure~\ref{fig3}). Since $R_m$ is the maximum range, the projectile can never reach beyond the enveloping parabola in any given direction $\alpha$. Therefore, the enveloping parabola delimits the region of the plane inaccessible to any projectile launched from the origin with speed $v$. In the literature this region is known as the {\it safe domain}~\cite{jean}. 

In passing, we note that students who are not familiar with the equation of parabola in polar coordinates may further transform (\ref{maxrange}) into the more familiar quadratic function as follows: rewrite (\ref{maxrange}) as $R_m(1+\sin\alpha) =v^2/g$, then transpose the term $R_m\sin\alpha = y_m$ and square both sides to find $x_m^2+y_m^2=v^4/g^2+2y_{m}v^2/g+y_m^2$, where use has been made of $R_m^2=x_m^2+y_m^2$. Solving the resulting equation for $y_m$, yields
\begin{equation}
\label{envpar}
y_m =-\frac{g}{2v^2}x_m^2 + \frac{v^2}{g}.
\end{equation}
This is the equation of the enveloping parabola (see figure~\ref{fig3}) in rectangular coordinates.

\section{Horizontal range with initial height}
\begin{figure}[H]
\centering
\includegraphics[scale=0.5]{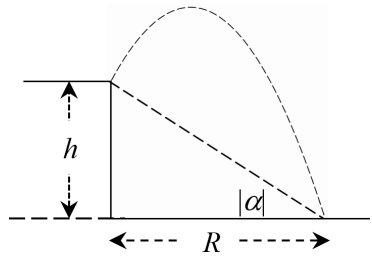}
\caption{A projectile launched from height $h$ above level ground.}
\label{fig4}
\end{figure}

Let us now consider the problem of maximizing the range of a projectile launched from a height $h$ above level ground (see figure~\ref{fig4}). In principle, we have already solved this problem! Using (9), one sets $y_m =-h$ and solves for $x_m$, which is the sought for maximum range. However, the instructor may choose to avoid mention of the enveloping parabola altogether. In this case, as we discuss below, travel times again provide a straightforward algebraic solution.

For a given range $R$, the launch angles are related as in the previous section, i.e. $\theta_1 + \theta_2 = \pi/2 - \alpha$,  where $\alpha$ (here negative) is such that $\tan\alpha=-h/R$.  As to the travel times, $t_1$ and $t_2$, they are the positive roots of $\frac{1}{4}g^2t^4-(v^2+gh)t^2+R^2+h^2=0$, and so satisfy the following relations
\begin{equation}
\label{vieta5}
t_1^2t_2^2=4\frac{R^2+h^2}{g^2},
\end{equation}
and
\begin{equation}
\label{vieta6}
t_1^2+t_2^2=4\frac{v^2+gh}{g^2}.
\end{equation}
Combining (\ref{vieta5}) and (\ref{vieta6}) gives
\begin{equation}
\label{range2}
\sqrt{R^2+h^2} = \frac{v^2+gh}{g} - \frac{1}{4}g(t_1-t_2)^2,
\end{equation}
from which it readily follows that the maximum range $R_m$, corresponding to $t_1 = t_2$, is given by
\begin{equation}
\label{maxrange2}
R_m=\frac{v^2}{g}\sqrt{1+\frac{2gh}{v^2}}.
\end{equation}
Although equations (\ref{vieta5}) and (\ref{vieta6}) easily lend themselves to a simple geometric interpretation, unlike the above cases, such an interpretation seems to be somewhat contrived as it involves the times $t_1$ and $t_2$ rather than the displacements. 

\section{Conclusion} 
We present a simple approach that employs travel times, and no calculus, to provide a simple solution to the maximum range problem for a projectile launched on level ground, with or without initial height, as well as a sloping ground with no initial height. For zero initial heights, our approach allows a transparent geometric interpretation within which the problem of maximizing the range receives an elegant geometric solution and the relation between the launch angles, for a given range $R$, obtains in a natural and visual way.

Moreover, our method can be used to devise simple constructive solutions to the range $R$ and travel time $t$ for any given initial speed $v$ and launch angle $\theta$. For example, using the given $v$, one calculates the diameter $2v^2/g$ and constructs the circle of figure~\ref{fig1}. Then using the given $\theta$, one locates on this circle the missing vertex and constructs a right-angled triangle with the diameter as hypotenuse and $\theta$ as an adjacent angle. Now $R$ and $vt$ are just the distance from the located vertex to the diameter and the side of the triangle opposite to $\theta$ respectively. Guided by figure~\ref{fig2}, one can easily conceive of a similar construction for the case of a sloping ground. The reader is invited to fill in the details.

In the paper we try to promote the non-calculus approach to tackle a set of problems on projectile motion, traditionally considered to be the privilege of calculus-based courses. It is hoped that this paper will make these issues, together with their non-calculus solutions, accessible to all introductory mechanics students regardless of their calculus erudition.

\end{document}